\documentstyle[11pt]{article}
\newcounter{subequation}[equation]

\makeatletter 
 
\def\thesubequation{\theequation\@alph\c@subequation}
\def\@subeqnnum{{\rm (\thesubequation)}}
\def\slabel#1{\@bsphack\if@filesw {\let\thepage\relax
   \xdef\@gtempa{\write\@auxout{\string
      \newlabel{#1}{{\thesubequation}{\thepage}}}}}\@gtempa
   \if@nobreak \ifvmode\nobreak\fi\fi\fi\@esphack}
\def\subeqnarray{\stepcounter{equation}
\let\@currentlabel=\theequation\global\c@subequation\@ne
\global\@eqnswtrue
\global\@eqcnt\z@\tabskip\@centering\let\\=\@subeqncr
$$\halign to \displaywidth\bgroup\@eqnsel\hskip\@centering
  $\displaystyle\tabskip\z@{##}$&\global\@eqcnt\@ne
  \hskip 2\arraycolsep \hfil${##}$\hfil
  &\global\@eqcnt\tw@ \hskip 2\arraycolsep
  $\displaystyle\tabskip\z@{##}$\hfil
   \tabskip\@centering&\llap{##}\tabskip\z@\cr}
\def\endsubeqnarray{\@@subeqncr\egroup
                     $$\global\@ignoretrue}
\def\@subeqncr{{\ifnum0=`}\fi\@ifstar{\global\@eqpen\@M
    \@ysubeqncr}{\global\@eqpen\interdisplaylinepenalty \@ysubeqncr}}
\def\@ysubeqncr{\@ifnextchar [{\@xsubeqncr}{\@xsubeqncr[\z@]}}
\def\@xsubeqncr[#1]{\ifnum0=`{\fi}\@@subeqncr
   \noalign{\penalty\@eqpen\vskip\jot\vskip #1\relax}}
\def\@@subeqncr{\let\@tempa\relax
    \ifcase\@eqcnt \def\@tempa{& & &}\or \def\@tempa{& &}
      \else \def\@tempa{&}\fi
     \@tempa \if@eqnsw\@subeqnnum\refstepcounter{subequation}\fi
     \global\@eqnswtrue\global\@eqcnt\z@\cr}
\let\@ssubeqncr=\@subeqncr
\@namedef{subeqnarray*}{\def\@subeqncr{\nonumber\@ssubeqncr}\subeqnarray}
\@namedef{endsubeqnarray*}{\global\advance\c@equation\m@ne%
                           \nonumber\endsubeqnarray}

\makeatletter
\@addtoreset{equation}{section}
\makeatother
\renewcommand{\theequation}{\thesection.\arabic{equation}}

\def\dalemb#1#2{{\vbox{\hrule height .#2pt
        \hbox{\vrule width.#2pt height#1pt \kern#1pt
                \vrule width.#2pt}
        \hrule height.#2pt}}}

\let\a=\alpha \let\b=\beta   \let\e=\epsilon
  \let\q=\theta  
  \let\n=\nu \let\x=\xi

\def\nn{\nonumber} \def\bd{\begin{document}} \def\ed{\end{document}}
\def\ds{\documentstyle} \let\fr=\frac \let\bl=\bigl \let\br=\bigr
\let\Br=\Bigr \let\Bl=\Bigl 
\let\bm=\bibitem
\let\na=\nabla
\let\pa=\partial \let\ov=\overline
\def\ie{{\it i.e.\ }} 
\newcommand{\be}{\begin{equation}} 
\newcommand{\ee}{\end{equation}} 
\def\ba{\begin{array}}
\def\ea{\end{array}}
\def\ft#1#2{{\textstyle{{\scriptstyle #1}\over {\scriptstyle #2}}}}
\def\fft#1#2{{#1 \over #2}}
\def\del{\partial}
\def\sst#1{{\scriptscriptstyle #1}}
\def\oneone{\rlap 1\mkern4mu{\rm l}}
\def\e7{E_{7(+7)}}
\def\td{\tilde}
\def\wtd{\widetilde}
\def\im{{\rm i}}
\def\bog{Bogomol'nyi\ }
\def\q{{\tilde q}}
\def\hast{{\hat\ast}}
\def\0{{\sst{(0)}}}
\def\1{{\sst{(1)}}}
\def\2{{\sst{(2)}}}
\def\3{{\sst{(3)}}}
\def\4{{\sst{(4)}}}
\def\5{{\sst{(5)}}}
\def\6{{\sst{(6)}}}
\def\7{{\sst{(7)}}}
\def\8{{\sst{(8)}}}
\def\n{{\sst{(n)}}}
\def\oo{{\"o}}
\def\hA{\hat{\cal A}}
\def\ns{{\sst {\rm NS}}}
\def\rr{{\sst {\rm RR}}}
\def\tH{{\widetilde H}}
\def\tB{{\widetilde B}}
\def\cA{{\cal A}}
\def\cF{{\cal F}}
\def\tF{{\wtd F}}
\def\Z{\rlap{\sf Z}\mkern3mu{\sf Z}}
\def\ep{{\epsilon}}
\def\IIA{{\rm IIA}}
\def\IIB{{\rm IIB}}
\def\ads{{\rm AdS}}
\def\R{\rlap{\rm I}\mkern3mu{\rm R}}

\def\Ei{{\hbox{Ei}}}
\def\Ci{{\hbox{Ci}}}
\def\Si{{\hbox{Si}}}

\newcommand{\ho}[1]{$\, ^{#1}$}
\newcommand{\hoch}[1]{$\, ^{#1}$}
\newcommand{\bea}{\begin{eqnarray}} 
\newcommand{\eea}{\end{eqnarray}} 
\newcommand{\ra}{\rightarrow}
\newcommand{\lra}{\longrightarrow}
\newcommand{\Lra}{\Leftrightarrow}
\newcommand{\ap}{\alpha^\prime}
\newcommand{\bp}{\tilde \beta^\prime}
\newcommand{\tr}{{\rm tr} }
\newcommand{\Tr}{{\rm Tr} } 
\newcommand{\NP}{Nucl. Phys. }
\newcommand{\tamphys}{\it Center for Theoretical Physics,
Texas A\&M University, College Station, TX 77843}
\newcommand{\upenn}{\it Dept. of Phys. and Astro., 
University of Pennsylvania,
Philadelphia, PA 19104}

\newcommand{\auth}{M. Cveti\v{c}\hoch{\dagger1}, H. L\"u\hoch{\dagger1}, 
C.N. Pope\hoch{\ddagger2} and T.A Tran\hoch{\ddagger}}

\begin{document}
\begin{flushright}
\hfill{CTP TAMU-05/99 \\ 
UPR/0830-T \\
January 1999}\\
\hfill{\bf hep-th/9901115}\\
\end{flushright}


\begin{center}
{\large {\bf Closed-form Absorption Probability of Certain $D=5$ and
$D=4$ Black Holes and Leading-Order Cross-Section of Generic Extremal
$p$-branes}}

\vspace{20pt}

\auth

\vspace{10pt}
{\hoch{\dagger}\upenn}

\vspace{10pt}
{\hoch{\ddagger}\tamphys}

\vspace{30pt}

\underline{ABSTRACT}
\end{center}

         We obtain the closed-form absorption probabilities for
minimally-coupled massless sca- lars propagating in the background of
$D=5$ single-charge and $D=4$ two-charge black holes.  These are the
only two examples of extremal black holes with non-vanishing
absorption probabilities that can be solved in closed form for
arbitrary incident frequencies.  In both cases, the absorption
probability vanishes when the frequency is below a certain threshold,
and we discuss the connection between this phenomenon and the
behaviour of geodesics in these black hole backgrounds.  We also
obtain leading-order absorption cross-sections for generic extremal
$p$-branes, and show that the expression for the cross-section as a
function of frequency coincides with the leading-order dependence of
the entropy on the temperature in the corresponding near-extremal
$p$-branes.

{\vfill\leftline{}\vfill
\vskip 10pt \footnoterule {\footnotesize \hoch{1} Research supported
in part by DOE grant DOE-FG02-95ER40893
\vskip  -12pt} \vskip   14pt
{\footnotesize
        \hoch{2}        Research supported in part by DOE 
grant DOE-FG03-95ER40917 \vskip -12pt}  \vskip  14pt
}

\pagebreak
\setcounter{page}{1}

\tableofcontents
\addtocontents{toc}{\protect\setcounter{tocdepth}{2}}
\newpage

\section{Introduction\label{sec:intro}}

       There has been considerable interest recently in studying
absorption probabilities for fields propagating in various black hole
and $p$-brane backgrounds\cite{dmw1,gk1,mast,cgkt,km,kleb,gkt,cl1,%
cl2,ghkk,lm,mm,gubhash,taylor,ah,clpt}.  One of the motivations is
inspired by the conjectured duality of supergravity on an AdS
spacetime and the conformal field theory on the boundary of the AdS
\cite{mald}.  To obtain the absorption probability is however by
itself a subject of interest.  The wave equation for a
minimally-coupled scalar in an extremal $p$-brane background depends
only on the metric of the $p$-brane, which has the form
\be
ds^2= \prod_{\a=1}^N H_\a^{-\ft{\td d}{D-2}}\, 
dx^\mu dx^\nu \eta_{\mu\nu}
+ \prod_{\a=1}^N H_\a^{\ft{d}{D-2}}\, dy^mdy^m
\ ,\label{dmetric}
\ee 
where $d=p+1$ is the dimension of the world volume of the $p$-brane
and $\td d =D-d-2$, and $H_\a= 1 + Q_\a/r^{\td d}$ are harmonic
functions in the transverse space $y^m$, with $r^2 =y^m\, y^m$.  It
follows that the wave equation $\del_{\sst M} (\sqrt{g}\,
g^{\sst{MN}}\del_{\sst N} \Phi)=0$ for the massless minimally-coupled
scalar, with the ansatz $\Phi(t, r, \theta_i) =\phi(r)\, Y(\theta_i)\,
e^{-\im \omega t}$, takes the following form:
\be
\fft{d^2\phi}{d\rho^2} + \fft{\td d+1}{\rho}\, \fft{d\phi}{d\rho} + 
\Big[
\prod_{\a=1}^N (1 +\fft{\lambda^{\td d}_\a}{\rho^{\td d}}) 
-\fft{\ell(\ell +\td d)}{\rho^2} \Big]\, 
\phi =0\ ,\label{genwave}
\ee
where $\rho=\omega\, r$ and $\lambda_\a = \omega\, Q_\a^{1/\td d}$.
Note that the wave equation depends on $\td d$, but is independent of
the world-volume dimension $d$.  This implies that the wave equation
is invariant under double dimensional reduction of the corresponding
$p$-brane.  It is not, however, invariant under vertical dimensional
reduction, where $\td d$ rather than $d$ is reduced.

    In general, the wave equation (\ref{genwave}) is not exactly
solvable; it cannot be mapped into any known second-order differential
equation.  The leading-order absorption probability can be obtained by
matching approximate solutions of the wave equation that are valid in
overlapping inner and outer regions; this technique was pioneered in
absorption probability calculations for the Schwarzschild black hole
\cite{unruh}.  Only for a few examples are the wave equations exactly
solvable, including the extremal D3-brane \cite{gubhash} and the
extremal dyonic string \cite{clpt}, corresponding to $N=1$, $\td d=4$
and $N=2$, $\td d =2$ respectively.  In these two cases, the wave
equation (\ref{genwave}) can be cast into the form of the modified
Mathieu equation \cite{gubhash,clpt}, which has been studied in great
detail in the mathematical literature \cite{doug}.  However, even in
these two cases the final results for the absorption probabilities can
only be expressed in terms of power-series expansions in some small
parameter, which depends on the frequency of the wave and, in the case
of dyonic string, on the relative ratio of the electric and magnetic
charges as well.

        In this paper, we shall consider two further examples, where
the absorption probability can in fact be obtained in closed form for
scalar waves of arbitrary frequencies.  One example is the extremal
single-charge black hole in $D=5$, and the other is the extremal
two-charge black hole in $D=4$.  We discuss these two examples in
sections 2 and 3 respectively.  In both cases, an interesting
phenomenon occurs, namely that the absorption probability vanishes
below a certain threshold frequency for the incident scalar wave.  In
section 4, we observe that this is related to special features of the
behaviour of infalling geodesics in these two black holes.

        Although the wave equation (\ref{genwave}) cannot be solved in
general, the leading-order absorption probability for low energy waves
can nevertheless be obtained by matching inner and outer solutions of
the wave equations.  We shall apply this technique in section 5, and
obtain the leading-order absorption for general extremal $p$-branes,
which we find to be given by
\be
\sigma = c\, (Q_1\, Q_2\, \cdots\, Q_N)^{\ft{\td d}{N\td d -2}}\,
\omega^{\ft{2\td d}{N\td d-2} -1}\ ,
\ee
where $Q_\a$ are the charges and $c$ is a charge-independent numerical
constant.  This result coincides with the leading-order dependence of
the entropy on the temperature of the
corresponding near-extremal $p$-brane, namely
\be
S=\td c\, (Q_1\, Q_2\, \cdots\, Q_N)^{\ft{\td d}{N\td d -2}}\,
T^{\ft{2\td d}{N\td d-2} -1}\ ,
\ee
which we obtain in section 6.  Such a parallel implies a connection
between the entropy of near-extremal $p$-branes and the low-energy
absorption by the corresponding extremal $p$-branes.

\section{Single-charge black hole in $D=5$}

     In maximal $D=5$ supergravity, each of the 27 vector potentials
can support a single-charge black hole.  These black holes form a
27-dimensional representation under the Weyl group of $E_6$
\cite{lpsweyl}.  A complete classification of the field configurations
that supports $N$-charge $p$-branes in maximal supergravities can be
found in \cite{classp}.  The single-charge black hole in $D=5$
corresponds to $N=1$, $\td d =2$, and hence from (\ref{genwave}) the
scalar wave equation is
\be
\fft{d^2\phi}{d\rho^2} + \fft3{\rho}\, \fft{d\phi}{d\rho} + 
\Big[ 1 + \fft{\lambda^2-\ell(\ell +2)}{\rho^2}\Big] \, \phi = 0
\ .\label{d5wave1}
\ee
This  can be solved exactly, giving
\be
\phi = \fft{\a}{\rho}\, J_{\im q}(\rho) + \fft{\beta}{\rho}\, 
J_{-\im q}(\rho)\ ,\label{d5singlesol}
\ee
where we have defined 
\be 
q \equiv \sqrt{\lambda^2 -(\ell +1)^2}\ .\label{d5qdef}
\ee
We see from this that the behaviour of the solutions will be radically
different depending on whether $\lambda$ is larger than $(\ell +1)$ or
smaller than $(\ell +1)$.  In particular, it will turn out that we
only have wave-like behaviour near the horizon if $\lambda > \ell +1$.
(It is always understood in our discussions that $\lambda$ is, without
loss of generality, taken to be non-negative.)

    Expanding (\ref{d5singlesol}) around $\rho=0$, using
(\ref{smallx}), we find that the wave-function near the horizon takes
the form
\be
\phi \sim \fft{\a}{\rho\, \Gamma(1+\im q)}\, e^{\im q\log(\rho/2)} 
+  \fft{\b}{\rho\, \Gamma(1-\im q)}\, e^{-\im q\log(\rho/2)}\ .
\label{d5nearhor}
\ee
(We have used the identity $x^y = e^{y\log x}$ here.) The boundary
condition at the horizon, which requires that the wave be purely
ingoing there, implies that we must have $\a=0$.

   Expanding (\ref{d5singlesol}) at large $\rho$, using
(\ref{largex}), we find that the wave-function near infinity takes the
form
\be
\phi \sim -\fft{\beta}{2\rho}\, \sqrt{\fft{2}{\pi\rho}}\, e^{\pi q/2}\,
e^{\im\pi/4}\, \Big( -e^{-\im\rho} + \im \, e^{-\pi q}\, e^{\im\rho}
\Big) \ .\label{d5infinity}
\ee
Thus comparing with  the generic structure $\phi \propto 
(-e^{-\im \rho} + S_0\, e^{\im \rho})$, we read off the S-matrix:
\be
S_0 = \im \, e^{-\pi q}\ .
\ee
From this, it follows that the absorption probability for the
single-charge $D=5$ extremal black hole is given by
\bea
P &=& 1 - |S_0|^2 = 1 -e^{-2\pi q} \nn\\
&=& 1-e^{-2\pi\sqrt{\lambda^2-(\ell +1)^2}}\ ,\qquad \lambda\ge \ell +
1\ .\label{d5absorb}
\eea
It should be emphasised that this is an exact result, correct to all
orders in $\lambda$.  It should also be noted that if $\lambda\le \ell
+1$ there is no wave-like behaviour near the horizon, and the
absorption probability is therefore zero for all $\lambda \le \ell
+1$.

     It is instructive also to compute the absorption probability in
this example by taking the ratio of the ingoing fluxes at the horizon
and at infinity.  The flux in this case is given in general by $F= \im
\rho^3\, (\bar\phi\, \del_\rho\phi - \phi\, \del_\rho\bar\phi)$, where
here $\phi$ is taken to be purely ingoing component of the scalar
wave.  Using the near-horizon form of the solution (\ref{d5nearhor})
(and recalling that the boundary condition required $\a=0$), we see
that flux into the horizon is given by
\be
F_{\rm horizon} = \fft{2|\b|^2\, q}{\Big| \Gamma(1-\im q)\Big|^2}
 = \fft{2|\b|^2}{\pi}\, \sinh\pi q \ .
\ee
On the other hand the ingoing flux at infinity, calculated using the
asymptotic form (\ref{d5infinity}), is
\be
F_\infty = \fft{|\b|^2}{\pi}\, e^{\pi q}\ .
\ee
Hence we find that the absorption probability is
\be
P = \fft{F_{\rm horizon}}{F_\infty} = 1 - e^{-2\pi q}\ ,
\ee
when $\lambda \ge \ell+1$, in agreement with (\ref{d5absorb}).

   Note that in the non-extremal case the wave equation is more
complicated.  The absorption probability for non-extremal 3-charge
black holes has been calculated for low energies in \cite{mast,km}, by
the standard techniques involving the overlap between approximate
solutions in inner and outer regions.  In particular, in \cite{km} two
of the three charges were taken to be small in comparison to third,
and can be taken to be zero.  In this case, when the frequency of the
$\ell$'th mode is such that $\lambda \le \ell+1$, the absorption can
be seen to be zero in the extremal limit, in agreement with our result
above.  On the other hand if $\lambda > \ell+1$, the approach in
\cite{km} yields an absorption probability for which the extremal
limit is singular, and it cannot be compared with the exact result
that we have obtained here.

\section{Two-charge black hole in $D=4$}

   Other examples where the wave equation is exactly solvable are for
the single-charge and two-charge extremal black holes in four
dimensions.  The single-charge case gives a wave equation which is
precisely equivalent to the familiar Coulomb problem in quantum
mechanics, and there is no absorption at any frequency \cite{balar}.
The two-charge case, on the other hand, is more analogous to the
single-charge $D=5$ black hole.  In this two-charge case, the wave
equation is
\be
\fft{d^2\phi}{d\rho^2} + \fft2{\rho}\, \fft{d\phi}{d\rho} + 
\Big[\Big(1 + \fft{\lambda_1}{\rho}\Big)
\Big(1 + \fft{\lambda_2}{\rho}\Big)-\fft{\ell(\ell +1)}{\rho^2}
\Big]\, \phi = 0\ ,\label{d4eqn}
\ee
where we now have two parameters, $\lambda_i = \omega\, R_i=\omega\
Q_i$, associated with the two charges $Q_1$ and $Q_2$.  (Note that
although the two-charge black hole can also be obtained from the
intersection of a D1-brane and a D5-brane by dimensional reduction, it
requires vertical as well as diagonal reduction steps.  Thus the wave
equation (\ref{d4eqn}) in $D=4$ is different from the D1-D5 wave
equation in $D=5$, 6 or higher dimensions, which can be cast into the
form of the modified Mathiew equation \cite{clpt}.) It is convenient
to define the following two constants:
\be
p\equiv \ft12(\lambda_1 + \lambda_2)\ ,\qquad q \equiv \sqrt{
4\lambda_1\, \lambda_2 -(2\ell +1)^2}\ .\label{pqdef}
\ee
The general solution to (\ref{d4eqn}), which is exact, is
\be
\phi = \a\, \rho^{(\im q-1)/2} \, e^{-\im\rho}\, U(\ft12 + \im p
+\ft{\im}{2} q, 1+\im q, 2\im\rho) +
  \b\, \rho^{(\im q-1)/2} \, e^{-\im\rho}\, M(\ft12 + \im p
+\ft{\im}{2} q, 1+\im q, 2\im\rho)\ ,\label{d4sol1}
\ee
where $U(a,b,z)$ and $M(a,b,z)$ are Kummer's irregular and regular
confluent hypergeometric functions, respectively.  We give some
details of their asymptotic behaviour in the Appendix.

     It can be seen from (\ref{mseries}) and (\ref{udef}) that the
behaviour of the exact wave function (\ref{d4sol1}) near $\rho=0$, in
the vicinity of the horizon, is given by
\bea
\phi&\sim &\fft{\im\,\a\pi \,  
\rho^{(\im q-1)/2}\, e^{-\im\rho}}{\sinh\pi q}
\, \Big\{ \fft1{\Gamma(\ft12+\im p -\ft{\im}{2} q) \, \Gamma(1+\im q)}
-\fft{(2\im)^{-\im q} \, \rho^{-\im q} }{
\Gamma(\ft12+\im p +\ft{\im}{2} q) \, \Gamma(1-\im q)} \Big\}\nn\\
&&+ \b\, \rho^{(\im q-1)/2}\, e^{-\im\rho}\ .\label{d4innersol}
\eea
Noting that we can write $\rho^{\pm\im q/2}$ as $e^{\pm\im
(q/2)\log\rho}$, we see that to satisfy the boundary condition that
the wave on the horizon be purely ingoing, the coefficient of
$\rho^{\im q/2}$ must vanish.  This gives the following relation
between $\a$ and $\b$:
\be
\beta = -\fft{\im\,\a\, \pi}{\Gamma(\ft12+\im p -\ft{\im}{2} q)\,
\Gamma(1+\im q)\, \sinh\pi q}\ .
\ee

    By looking at the large-$\rho$ asymptotic expansions for the Kummer
confluent hypergeometric functions, given by (\ref{muasymp}), we find
that the wave function near infinity can be written in the form
\be
\phi \sim - \fft{\a\, e^{\im\, \pi/4}\,
2^{-\im p-\im q/2}\, e^{\pi(p/2+q/4)}\, 
(e^{\pi q}+ e^{-2\pi p})\, \rho^{-\im p}}{
2\sqrt2\, \sinh(\pi q)\, \rho}\,  
\Big( -e^{-\im\rho} + S_0\, e^{\im\rho} \Big) \ ,
\ee
with the S-matrix $S_0$ given by
\be
S_0=\fft{2\, \im\, \Gamma(\ft12-\im p+\ft{\im}{2} q)\, (2\rho)^{2\im
p}\, \,e^{-\pi p}\,
\cosh\pi(p-\ft12 q)}{ \Gamma(\ft12+\im p+\ft{\im}{2} q)(e^{\pi q} 
+ e^{-2\pi p})}\ .
\ee
Consequently, we find that the absorption probability for the extremal
two-charge black hole in four dimensions is given by
\be P = 1 -|S_0|^2 = \fft{1-e^{-2\pi q}}{1 + e^{-\pi(q+2p)}}
\ ,\label{d4pcalc} \ee
when $\lambda_1\, \lambda_2 \ge (\ell+\ft12)^2$, while $P=0$ otherwise.
(In deriving these formulae, we have used the fact that the Gamma
function obeys the identity $|\Gamma(\ft12 + \im \, x)|^2 = \pi\,
{\rm sech}(\pi\, x)$, where $x$ is real.)  In terms of the original
dimensionless parameters $\lambda_i = \omega\, Q_i$, the absorption
probability is therefore given by
\be
P = \fft{1-e^{-2\pi\sqrt{4\lambda_1\, \lambda_2 -(2\ell +1)^2}}}{
1+ e^{-\pi(\lambda_1 +\lambda_2 +\sqrt{4\lambda_1\, \lambda_2 -
(2\ell+1)^2})}}\ , \qquad \lambda_1\, \lambda_2 \ge (\ell+\ft12)^2\ ,
\label{d42chargeabs}
\ee
with $P=0$ if $\lambda_1\, \lambda_2 \le (\ell+\ft12)^2$.

    Again, we may verify that the same result for the absorption
probability is obtained by calculating the ratios of the ingoing
fluxes at the horizon and at infinity.  In this case, the flux is
given by $F=\im\, \rho^2\, (\bar\phi\,\del_\rho\phi -\phi\,
\del_\rho\bar\phi)$.  From the results given above we find, after
simple algebra, that the ingoing fluxes are given by
\be
F_{\rm horizon}= \fft{|\a|^2\, e^{\pi q}\, \cosh\pi(p+\ft12q)}{
\sinh(\pi q)}\ ,\qquad
F_\infty = \fft{|\a|^2\, e^{\pi(p+q/2)}\, (e^{\pi q} + e^{-2\pi
p})^2}{4\sinh^2(\pi q)}\ .
\ee
Taking the ratio $F_{\rm horizon}/F_\infty$, we indeed find that it
agrees precisely with (\ref{d42chargeabs}).

    An alternative way of performing the calculation here is to
redefine the notion of what constitutes an ingoing wave and what
constitutes an outgoing wave.  This can be done, for example, simply
by defining the time dependence to be $e^{\im\omega t}$ rather than
$e^{-\im\omega t}$.  Since $\omega$ appears quadratically in the
original wave equation, this means that we can simply choose
universally to interpret the direction of motion of a wave front to be
the opposite of the ``conventional'' choice.  An alternative way of
expressing this is that since the wave equation such as (\ref{d4eqn})
for an energy eigenstate is real, one can complex conjugate the
solution (\ref{d4sol1}) and obtain another solution.

    The upshot is that we can choose to reinterpret the requirement
that the wave described by (\ref{d4sol1}) be ingoing at the horizon as
the requirement that the coefficient of $e^{-\im q/2}$ in
(\ref{d4innersol}) vanish, provided that we also define an ingoing
wave at infinity to be one with $e^{\im \rho}$, as opposed to
$e^{-\im\rho}$, dependence.  If we do this, then the requirement that
the wave on the horizon be ingoing becomes the condition $\a=0$, so
that the solution is
\be
\phi= \b\, \rho^{(\im q-1)/2} \, e^{-\im\rho}\, M(\ft12 + \im p
+\ft{\im}{2} q, 1+\im q, 2\im\rho)\ .\label{d4sol2}
\ee
This is a simpler form for the solution than the one we discussed
previously.  At large $\rho$, (\ref{d4sol2}) is proportional to
$(-e^{\im\rho} + S_0\, e^{-\im\rho})$, with
\be
S_0 = \fft{\im\, e^{-\pi q/2}\, (2\rho)^{-2\im\, p}\, \Gamma(\ft12
+\im\, p +\fft{\im}{2}\, q)}{\Gamma(\ft12 -\im\, p + \ft{\im}{2}\,
q)}\ .\label{d4s02}
\ee
It is easy to see that this gives rise to the same result
(\ref{d4pcalc}) for the absorption probability.

  In \cite{balar}, it was observed that the solution (\ref{d4sol2})
could be expressed at large $\rho$ in the form
\be
\phi \sim \fft{1}{\rho}\, \sin \Big(\rho + p\, \log(2\rho)
 - \ft12 L\, \pi  +\delta_{\sst L}\Big)\ ,
\ee
where $L=-\ft12 +\ft{\im}{2}\, q$ (we have adjusted the notation of
\cite{balar} to fit in with ours).  It follows from (\ref{d4s02}) that
$\delta_{\sst L}$ is a complex quantity, which we find to be
\be
e^{2\im\, \delta_{\sst L}} =\fft{\Gamma(\ft12 -\im\, p +\ft{\im}{2}\,
q)}{\Gamma(\ft12 +\im\, p +\ft{\im}{2}\, q)}\ .
\ee
(Note that unlike the standard situation for Coulomb scattering, or
single-charge black holes in $D=4$, where $\delta_{\sst L}$ is a real
quantity given by the phase of the relevant Gamma function, here
$\delta_{\sst L}$ is not equal to the real quantity ${\rm arg}\,
\Gamma(\ft12 -\im\, p +\ft{\im}{2}\, q)$, owing to the presence of the
$q$ term.) If the frequency in the $\ell$'th mode is such that
$\lambda_1\, \lambda_2 \le (\ell+\ft12)^2$, then both $L$ and
$\delta_{\sst L}$ are real and there is no absorption, while if
$\lambda_1\, \lambda_2 > (\ell+\ft12)^2$ they are both complex, and
the absorption is given by (\ref{d4pcalc}), with $p$ and $q$ given by
(\ref{pqdef}).

\section{Geodesics and absorption probability}

      In the previous sections, we have obtained exact absorption
probabilities in closed form for the extremal $D=5$ single-charge and
$D=4$ two-charge black holes.  The results are valid for massless
scalar waves of arbitrary frequency.  These two examples are of
interest since such exact solutions for absorption probabilities are
rare in gravity theories.  In both cases, the results exhibit an
interesting phenomenon, namely that the absorption probability is zero
in a given mode if the frequency of the scalar wave is below some
threshold value, related to the angular momentum $\ell$ of the mode.
A related, but more extreme, situation also occurs in the $D=4$
single-charge black hole, where the absorption probability vanishes
for all values of the frequency of the scalar wave.

        The vanishing of the absorption probabilities below certain
threshold frequencies in the two cases studied in this paper is
related to the behaviour of geodesics in these black holes.  To see
this, let us first consider radially-infalling timelike geodesics in
the metric of an extremal $N$-charge $p$-brane in $D$ dimensions.
These are described by the Lagrangian
\be
L = \ft12 g_{\mu\nu}\, \fft{dx^\mu}{d\tau}\, \fft{dx^\nu}{d\tau}
= -\ft12 \prod_\a H_\a^{-\fft{\td d}{D-2}}\, 
\Big(\fft{dt}{d\tau}\Big)^2 +
\ft12 \prod_\a H_\a^{\fft{d}{D-2}}\, 
\Big(\fft{dr}{d\tau}\Big)^2 \ ,
\ee
where $d=p+1$ and $\td d=D-d-1$.  The Lagrangian takes the value
$L=-\ft12$ for a timelike geodesic and so this, together with the
equation of motion for $t$, immediately gives us two first integrals:
\bea
\fft{dt}{d\tau} &=& E\, \prod_\a H^{\fft{\td d}{D-2}}\ ,\nn\\
\Big(\fft{dr}{d\tau}\Big)^2  &=& E^2\, 
\prod_\a H^{\fft{\td d -d}{D-2}} -
\prod_\a H^{-\fft{d}{D-2}}\ ,
\eea
where $E$ is a constant of integration.  Thus near the horizon at
$r=0$, we have that $H_\a \sim r^{-\td d}$, and hence
\be
\fft{dr}{dt} = \fft{dr}{d\tau}\, \fft{d\tau}{dt} \sim - 
r^{\fft12 N\, \td d}\ .
\ee
Thus we see that the coordinate time $t$ taken for the geodesic to
reach the horizon is either finite, logarithmically-divergent, or
power-law divergent as a function of $r$, according to the following
inequality:
\bea
\ft12 N\, \td d < 1:&&\qquad 
\hbox{Finite coordinate time} \nn\\
\ft12 N \, \td d = 1:&&\qquad 
\hbox{Logarithmically-divergent coordinate time} \label{cond}\\
\ft12 N \, \td d  > 1:&&\qquad 
\hbox{Power-law-divergent coordinate time} \nn
\eea
It is straightforward to verify that the same conclusions hold for
null geodesics.  (Note that in the case of black holes we have $\td
d=D-3$, but that the conditions (\ref{cond}) apply equally to any
extremal $N$-charge $p$-branes.)  Thus we see that amongst the
single-charge black holes, the case $D=4$ is distinguished by the fact
that timelike or null geodesics reach the horizon in a finite
coordinate time. (See, for example, \cite{cd}.) In $D=5$, the time
taken is logarithmically divergent, whilst in $D\ge 6$ it is power-law
divergent.  The 2-charge black hole in $D=4$ again has a
logarithmically-divergent coordinate time, while for all $N\ge3$
charge black holes in $D\ge 4$, and all $N\ge2$ charge black holes in
$D\ge5$, it again takes a power-law-divergent coordinate time to reach
the horizon.

   We therefore see the following correspondences.  If a timelike or
null geodesic reaches the horizon in finite coordinate time, there is
no absorption at all.  If it takes a logarithmically-divergent
coordinate time, there is absorption provided that the frequency of
the ingoing wave exceeds some finite bound, $\omega>\omega_0$; this is
the situation for the $D=5$ single-charge, and $D=4$ two-charge black
holes that we have considered in this paper.  Finally, if a timelike
geodesic reaches the horizon in a power-law-divergent coordinate time,
there is non-vanishing absorption for all frequencies $\omega>0$.

       The correspondence between the behaviour of infalling geodesics
and the absorption probability may not be too surprising, in view of
the fact that the radial coordinate-dependence of the scalar wave
function near the horizon of a $p$-brane is closely related to the
dependence of the radial coordinate of an infalling geodesic on the
coordinate time $t$.  For example, the coordinate time $t$ depends
logarithmically on the radial coordinate $r$ of an infalling geodesic
in the $D=5$ single-charge and $D=4$ two-charge black holes.  A
similar dependence, namely $\phi\sim e^{\im k\log\rho}$, appears in
the near-horizon form of the solutions (\ref{d5nearhor},
\ref{d4innersol}) of the wave equation in the corresponding black-hole
backgrounds.  More generally, for any extremal $p$-brane the wave
equation is given by (\ref{genwave}), and the solution near the
horizon at $\rho=0$ is given by (\ref{insol}).  The near-horizon
solution is therefore wavelike if $\mu\ge0$, but instead has just a
power-law dependence on $\rho$ if $\mu<0$.  The sign of $\mu$, which
is given by (\ref{munu}), is thus precisely correlated with the cases
enumerated in (\ref{cond}).

\section{Leading order absorption by extremal $p$-branes}

       In general the wave equation (\ref{genwave}) cannot be solved
analytically.  One approach is to solve the equation in two distinct
but overlapping regions, namely an inner region near to the $p$-brane,
and an outer region reaching to infinity.  Such a technique has been
widely used recently in obtaining low energy absorption sections for
various black holes and $p$-branes \cite{dmw1,gk1,mast,cgkt,km,%
cl1,cl2,kleb,gkt,ghkk,lm,mm,taylor,ah}.

          For the outer solution, it is convenient to define $\phi
=\rho^{-(\td d +1)/2}\, \psi$, leading to the wave equation
\be
\psi'' + \Big[ \fft{1-(2\ell +\td d)^2}{4\rho^2}
+ \prod_{\a=1}^N \Big( 1 + \fft{\lambda_\a^{\td d}}{\rho^{\td d}}\Big)
\Big]\psi =0\ .\label{outeq}
\ee
For notational convenience, we define
\be
\Lambda = (\prod_{\a=1}^N \lambda_\a)^{1/N}\ .
\ee
(This means that if all the charges are equal we have
$\lambda_\a=\Lambda$.) In the region where
\be
\rho >> \Lambda^{N\td d/(N \td d -2)}\ ,\label{outregion}
\ee
the equation (\ref{outeq}) can be approximated as
\be
\psi'' + \Big[ \fft{1 - (2\ell +\td d)^2}{4\rho^2} + 1
\Big] \psi =0\ ,
\ee
which can be solved in terms of  Bessel functions.   Thus the outer
solution is given by
\be
\phi_{\rm outer} = \rho^{-\td d/2}\Big(\alpha\, 
J_{\ell +\td d/2}(\rho) +
\beta\, Y_{\ell +\td d/2}(\rho) \Big)\ .\label{outsol}
\ee
Note that the constraint (\ref{outregion}) is derived with the
assumption that $\td d >2$.  Otherwise the constraint would be $\rho
>> \Lambda$.  Thus the cases with $\td d =2$ and 1 are
exceptional; these were discussed in the previous sections.

       In the inner region, the constant 1 in the harmonic function
can be dropped, provided that \footnote{In the case of $N=1$, the
valid region can actually be $\rho <<1$.}
\be
\rho << \Lambda\ .\label{inregion}
\ee
It is convenient to use a new variable $z$, and to define a new
wavefunction, as follows:
\be
z = \fft{\Lambda^{N\td d/2}}{u\,\rho^{u}}\ ,\qquad
\phi = z^{\ft12(v - 1)}\, f(z)\ ,\label{1zdef}
\ee
where 
\be
u = (N\td d -2)/2\ ,\qquad v = 2\td d/(N\td d-2)\ .\label{munu}
\ee
The equation (\ref{genwave}) then becomes
\be
f'' + \Big[ \fft{1-(2L +v)^2}{4\rho^2}
+ \prod_{\a=1}^N \Big( 1 + \Big(\fft{\Lambda}{\lambda_\a}\Big)^{\td d}\, 
\fft{(\Lambda/u)^{v}}{z^{v}}\Big)
\Big]\,f =0\ .\label{ineq}
\ee 
where $L=2\ell/(N\td d -2)$.  Specifically, when $N=1$ we have $v >
2$, and hence the limit where the approximation to (\ref{ineq}) is a
good one is when $z >> \Lambda^{N\,v/(N\,v-2)}$, which implies $\rho
<< 1$.  If, on the other hand we have $N>1$, then $v < 2$, and hence
the approximation is good when $z>> \Lambda$, implying
(\ref{inregion}).  If we make this approximation then equation
(\ref{ineq}) becomes
\be
f'' + \Big[ \fft{1-(2L +v)^2}{4 z^2} + 1 
\Big]\,f =0\ .\label{ineq2}
\ee 
which is solvable in terms of Bessel functions, giving
\bea
\phi_{\rm inner} &=& z^{v/2}\, (c_1\, J_{L +v/2}(z) +
c_2\, Y_{L +v/2}(z)) \nn\\
&=&\fft{\Lambda^{v N \td d/4}}{u^{v/2}\, \rho^{\td d/2}}\,
\Big[ c_1\, J_{L +v/2}(\fft{\Lambda^{N\td d/2}}{u\,\rho^{u}}) +
c_2\, Y_{L +v/2}(\fft{\Lambda^{N\td d/2}}{u\,\rho^{u}})
\Big]\ .\label{insol}
\eea
In order for the solution to describe a purely ingoing wave near the
horizon at $\rho=0$, we see from (\ref{largex}) that we must have $c_2
= \im\, c_1$.  The ingoing wave on the horizon is then given by
$\phi\sim -\im\, c_2\,\sqrt{2/\pi} \, z^{(v-1)/2}\, e^{\im\, z}\,
e^{-\im \,\pi(2L +v +1)/4}$.

        The inner and outer solutions are valid in the
regions specified in (\ref{inregion}) and
(\ref{outregion}), and thus it follows that in order to have an
overlap where there is a range of common validity for the two
solutions, the frequency $\omega$ must be sufficiently small that
\be
\fft{\Lambda^{N\,v}}{\Lambda} = \Lambda^{1/u} << 1
\ee
if $N>1$, while instead $\Lambda^{N\,v} <<1$ if $N=1$. This implies
that the approximations are valid in an appropriately low-energy
regime.  In such a regime, both the coordinates $z$ and $\rho$,
appearing respectively in the Bessel functions in the inner and outer
regions, are much less than 1, and so we can perform small-argument
series expansions on the Bessel functions in both regions, and hence
we can match the two solutions in the leading orders.  In the inner
region the boundary condition on the horizon has already determined
that the integration constants $c_1$ and $c_2$ are of equal modulus,
since $c_2=\im\, c_1$, but in the outer region the relative sizes of
the integration constants $\a$ and $\beta$ are yet to be determined.
However, since the structure in the inner region is established, it is
a simple matter to recognise the most dominant terms in the small-$z$
power-series expansion of $\phi_{\rm inner}$, and then to require that
these match with the small-$\rho$ power-series expansion of $\phi_{\rm
outer}$.  Note that we need to match both the functions and their
derivatives in the overlap region, implying that the functional forms
of the inner and outer solutions at leading order must be identical.
Structurally, these have the following forms:
\bea
{\rm Inner}:&& \phi\sim c_1\, z^{v+ L} + c_2\, 
(z^{-L} + z^{-L +2})\ ,\nn\\
&& \quad \sim c_1\, \rho^{-\td d-\ell} + c_2\, (\rho^{\ell} 
+ \rho^{\ell +2u})
\ , \nn\\
{\rm Outer}:&& \phi\, \sim \alpha (\rho^\ell + \rho^{\ell +2}) + 
\beta\, \rho^{-\td d-\ell}
\ .
\eea
It is straightforward to see that $z^{-L}$, (or equivalently
$\rho^{\ell}$), is the leading order of the expansion, since $z<<1$ in
the overlapping region.  For generic value of $L=2\ell/(N\td d -2)$,
the subleading order term is $z^{-L +2}$.  It follows that in general
we can only match the leading order term, providing a determination of
$\a$, but not $\beta$.  However, in some special cases, namely $v + L
< 2 -L$, {\it i.e.}
\be
\fft{2\ell + \td d}{N\td d -2} < 1\ ,\label{condxx}
\ee
the subleading order term is $z^{v +L}$.  In these cases, we can match
the inner and outer solutions at the level both of the leading
constant order and also the first sub-leading order $\rho^{-\td d}$.
Equating the coefficients of $\rho^\ell$ and $\rho^{-\td d-\ell}$ in
the overlap region thus enables us to obtain leading-order expressions
for both the integration constants $\a$ and $\beta$ in the outer
solution.  Note that for $\ell =0$, the condition (\ref{condxx}) is
satisfied provided that $N\ge2$.

    There are two methods available for calculating the absorption
probability.  The first is by calculating the ratio of the ingoing
flux at the horizon, divided by the ingoing flux at infinity.  This
method does not require the knowledge of the constant $\beta$.  The
second is by calculating the S-matrix, which is governed by the ratio
of $\beta$ to $\alpha$.  This second method can be used only if
$\beta$, in addition to $\alpha$, is known.  For now, we shall
therefore employ the flux-ratio method.

     As we saw above, to determine $\a$ we need only to keep
the leading order in a power-series expansion.   We find from
(\ref{insol}) and (\ref{smallx}) that this is given by
\be
\phi_{\rm inner} = -\fft{c_2}{\pi}\, 2^{L +v/2}\, 
\fft{u^L}{\Lambda^{N\td d L/2}}\, \Gamma(L +v/2)\, \rho^{\ell} 
+\cdots \ .\label{innerconst}
\ee
On the other hand for the outer solution, we find from (\ref{outsol})
that in the small-argument expansion appropriate to the overlap region
the leading-order term is given by
\be
\phi_{\rm outer} \sim \fft{\a\, 2^{-\ell -\td d/2}}{\Gamma(\ell + 
1+\ft12 \td d)} \, \rho^{\ell} +\cdots \ .
\label{outerconst}
\ee
(The $\b$ coefficient in the outer solution is negligible, at this
leading order, in comparison to $\a$.)  Equating the expressions in
(\ref{innerconst}) and (\ref{outerconst}), we obtain the relation
between the coefficients $\a$ and $c_2$ of the outer and inner
solutions. The absorption coefficient is most easily calculated at
this leading order as the ratio of the ingoing flux at the horizon,
divided by the ingoing flux at infinity.  In general, this flux may be
defined as
\be
F= \im\, \rho^{\td d+1}\, \Big(\bar\phi\, \fft{\del\phi}{\del\rho} -
\phi\, \fft{\del\bar\phi}{\del\rho} \Big)\ ,\label{flux}
\ee
where $\phi$ here is taken to be the ingoing component of the wave.
From the asymptotic forms of $\phi_{\rm inner}$ and $\phi_{\rm outer}$
where the arguments of the Bessel functions are large, we find from
(\ref{insol}), (\ref{outsol}) and (\ref{largex}) that the
ingoing fluxes at the horizon and at infinity are given by
\be
F_{\rm horizon} = \fft{4}{\pi} \, |c_2|^2\, u^{1-v}\,
\Lambda^{N\td d\,v/2}\ ,\qquad
F_\infty = \fft{|\a|^2}{\pi}\ ,\label{horinfflux}
\ee
and hence to leading order the absorption probability $P\equiv F_{\rm
horizon}/ F_\infty$ is
\be
P_{\ell} =
\fft{2 \pi^2\,  (2u)^{1-v -2L}\, \Lambda^{N\td d(L +v/2)}}{
2^{2\ell + \td d}\, \Gamma(\ell +1 + \ft12 \td d)^2\, 
\Gamma(L+ \ft12 v)^2}\ .
\ee
Finally, we note that the phase-space factor relating the absorption
probability to the scattering cross-section $\sigma$ is \cite{gubser}
\be
\sigma_\ell = 2^{\td d} \, \pi^{\td d/2}\, 
\Gamma(\ft12 \td d)\, \Gamma(\ell + 1+\ft12\td d)\, 
{\ell + \td d -1 \choose \ell}\, \omega^{-2\ell -\td d-1}\, P_\ell\ .
\ee
Hence we arrive at the result for the scattering cross-section
\be
\sigma_\ell =
\fft{2\pi^{2 + \td d/2}\, (2u)^{1-v-2L}\, \Gamma(\td d/2)\, 
\Lambda^{N\td d(L+\ft12 v)}\, \omega^{-\td d-1}}{
2^{2\ell}\, \Gamma(\ell + 1 +\ft12 \td d)\, \Gamma(L + \ft12 v)^2}\, 
{\ell +\td d -1 \choose \ell}\ .
\ee

          The above result is the leading-order contribution to the
low-energy absorption.   In this case the $\ell=0$ s-wave absorption
is dominant, and is given by
\be
\sigma_{0} =
\fft{2 \pi^{2 + \td d/2}\, (2u)^{1-v}\, \Gamma(\ft12 \td d)\, 
(Q_1Q_2\cdots Q_n)^{v/2} \omega^{v -1}}{
\Gamma( 1+ \td d ) \, \Gamma(\ft12 v)^2}\ .
\ee
Thus at low energy the absorption cross-section $\sigma$ of a generic
extremal $p$-brane, and the frequency of the incoming wave, are
related by
\be
\sigma = c\, (Q_1\, Q_2\, \cdots\, Q_N)^{\ft{\td d}{N\td d-2}}\, 
\omega^{\ft{2\td d}{N\td d -2} -1}\ ,\label{so}
\ee
where $c$ is some purely numerical constant.  It should be recalled
that the above result is applicable only when $N\td d -2>0$.  If
instead $N\td d-2\le 0$, which arises for $D=5$ single-charge and
$D=4$ single or two-charge black holes, the absorption probabilities
vanish at the low energies, below certain thresholds.  The closed-form
results in these cases were obtained in sections 2 and 3. Note that
when the condition (\ref{condxx}) is satisfied, (for example, $\ell=0$
and $N\ge 2$) both $\a$ and $\beta$ in (\ref{outsol}) can be
determined by matching the inner and outer solutions in the overlap
region.  In this case, the absorption can also be calculated by the
S-matrix method.

        In the next section we shall show that the leading-order
relationship (\ref{s0}) between the cross-section and the frequency
coincides with the leading-order relation between the entropy and the
temperature of the corresponding near-extremal $p$-brane.

      Finally, in this section, we note that the original wave
equation (\ref{outeq}), and the transformed wave equation
(\ref{ineq}), are identical in form if we make the replacements
\be
\td d \longleftrightarrow v\, \qquad
\ell \longleftrightarrow L\ ,
\qquad \lambda_\a \longleftrightarrow  \fft{\Lambda}{u}\,
(\fft{\Lambda}{\lambda_\a})^{\td d/v}\ .
\ee
In cases where $v$ is an integer, and furthermore is equal to $ d$,
then $(\td d, d)$ form a dual pair, associated with $(d-1)$-branes and
$(\td d-1)$-branes in $D=d + \td d +2$ dimensions.  When such a dual
pair arises, the dilaton becomes constant near the horizon of the
$p$-brane.  In cases such as the D3-brane, the dyonic string or the
$D=4$ four-charge black hole, we have $\td d=d$ and $L=\ell$, and in
these cases it follows that there is an exact duality symmetry of the
wave equation in the inner region and the outer region, for arbitrary
partial waves.  On the other hand for M2-brane and M5-brane, the
duality that maps between the inner and outer solutions is exact only
for the s-wave, as was observed in \cite{ghkk}.  For example, the
outer solution in the M2-brane with partial wave-number $\ell$ is
mapped to the inner solution in the M5-brane with partial wave-number
$\ell/2$.  The case for the three-charge $D=5$ black-hole/string
duality is analogous.  The outer solution in the black hole background
with partial wave-number $\ell$ is mapped to the inner solution in the
string background with partial wave-number $\ell/2$.  This occurrence
of half-integers may be indicative of an intrinsic r\^ole for fermions
in this discussion of duality.

\section{$S(T)$ {\it versus} $\sigma(\omega)$}

      The extremal $p$-brane solutions (\ref{dmetric}) can easily be
generalised to non-extremal ones, with \cite{dlpblack,ct}
\bea
ds^2 &=& \prod_{\a=1}^N H_\a^{-\ft{\td d}{D-2}}\, (-e^{2f}\, dt^2 +
dx^i\, dx^i) + \prod_{\a=1}^N H_\a^{\ft{d}{D-2}}\, (e^{-2f}\, dt^2 + 
r^2\, d\Omega^2)\ ,\nn\\
e^{2f} &=& 1 -\fft{k}{r^{\td d}}\ ,\qquad
H_\a = 1 + \fft{k}{r^{\td d}}\, \sinh^2\mu_\a\ .\label{nonextmetric}
\eea
The $N$ charges $Q_\a$ are given in terms of the parameters $k$ and
$\mu_\a$ by
\be
Q_\a = \ft12k\, \sinh 2\mu_\a\ .\label{charge}
\ee
The extremal limit is obtained by sending $k\longrightarrow 0$ and
$\mu_\a \longrightarrow \infty$, while keeping the $Q_\a$ fixed.  

         The outer horizon of the metric (\ref{nonextmetric}) is
located at $r_+ = k^{1/\td d}$.  It straightforward to show that the
Hawking temperature and entropy are given by
\be
T =\fft{\td d}{4\pi r_+}\, \prod_{\a=1}^N (\cosh\mu_\a)^{-1}\ ,\qquad
S = \ft14 r_+^{\td d +1}\, \Omega_{\td d+1}\, \prod_{\a=1}^N
\cosh\mu_\a\ ,
\ee
where $\Omega_n$ is the volume of the unit $n$-sphere.  For
convenience, we may define a ``scaled'' entropy $\wtd S \equiv
(16\pi/(\td d\, \Omega_{\td d +1})\, S$, in terms of which we have
\be
\wtd S\, T = k\ .
\ee
 From the expression (\ref{charge}) for the charges,  we see that
\be
\cosh\mu_\a = \sqrt{\fft{k + \sqrt{k^2 + 4 Q_\a^2}}{2k}}\ .
\ee
From this, it follows that
\bea
\wtd S &=& \fft{4\pi}{\td d}\, k^{1-1/v}\, \prod_{\a=1}^N
\sqrt{\ft12 k + \sqrt{\ft14 k^2 + Q_\a^2}}\nn\\
&=& \fft{4 \pi}{\td d}\, (\wtd S\, T)^{1 -1/v}\,
\prod_{\a=1}^N \sqrt{\ft12 (\wtd S\, T) +
\sqrt{\ft14 (\wtd S\, T)^2 + Q_\a^2}}\ ,
\eea
where $v$ is defined in (\ref{munu}).  Hence we obtain an implicit
relationship between entropy and temperature, given by
\be
\wtd S = \td c\, T^{v -1}\, 
\prod_{\a=1}^N \Big(\ft12 (\wtd S\, T) + \sqrt{\ft14 (\wtd S\, T)^2 +
Q_\a^2}\Big)^{v/2}\ ,\label{strecursion}
\ee
where we have defined $\td c = (\fft{4\pi}{\td d})^{v}$.

    In the near-extremal regime, where $\wtd S\, T=k <<Q_\a$, we
can therefore obtain the leading-order expansion of entropy in terms of
temperature, given by
\be 
\wtd S_0 = \td c\, (Q_1\, Q_2\, \cdots\, Q_N)^{\ft{\td d}{N\td d -2}}
\,\, T^{\ft{2\td d}{N\td d -2} -1}\ .  \label{s0}
\ee
Thus the leading-order relationship between the entropy and
temperature of a near-extremal $p$-brane coincides precisely with the
leading-order relationship (\ref{so}) between the scattering
cross-section and the frequency for scalar s-waves in the
corresponding extremal $p$-brane.  (Such a correspondence between
$S(T)$ and $\sigma(\omega)$ breaks down when $N\td d -2 \le 2$, which
is associated with finite or logarithmically-divergent coordinate time
for geodesics to reach the horizon.  In these cases, the temperature
increases as the $p$-brane approaches extremality.  When $N\td d=2$,
the temperature reaches a finite maximum at the extremal limit, whilst
for $N\td d <2$, the temperature goes to infinity at the extremal
limit.  Such black holes were discussed in \cite{holwil}, and they can
be viewed as elementary particles.)

       It is of interest to look at the higher-order terms in the
expansion for the entropy as a function of temperature.  Note that the
expansion parameter is $k=(\wtd S_0\, T) \sim T^{v}$.  The expansion
can easily be obtained from the implicit expression for $S$ in terms
of $T$ given in (\ref{strecursion}), by means of the following
iterative algorithm.  If we define $\wtd S_i$ to be the entropy
accurate up to order $T^{(i+1)v -1}$, then by substuting $\wtd S_i$
into the right-hand side of equation (\ref{strecursion}), we obtain a
new expression $\wtd S_{i+1}$ that is accurate up to order
$T^{(i+2)v -1}$:
\be
\wtd S_{i+1} = \td c\, T^{v -1}\, 
\prod_{\a=1}^N \Big(\ft12 (\wtd S_i\, T) + \sqrt{\ft14 (\wtd S_i\, T)^2 +
Q_\a^2}\Big)^{v/2}\ ,\label{strecursion2}
\ee
With the leading order $\wtd S_0$ given by (\ref{s0}), we can then
obtain the relation between the entropy and the temperature to
arbitrary order, as a power series in $T^v$.  We find that the first
few orders are
\bea
\fft{S}{S_0}&=&  1 + \ft14 \xi_1\, T^v + \ft3{32} \xi_1^2\,
T^{2v} +
\ft{1}{96}(4\xi_1^3 -\xi_3)\, T^{3v} + 
\ft{5}{6144}(25 \xi_1^4 - 16 \x_1\, \xi_3)\, T^{4v} \nn\\
&&+
\ft{3}{2560} (\x_1^5 - 10 \x_1^2\, \x_3 + \x_5)\, T^{5v}+
{\cal O}(T^{6v})\ ,\label{st}
\eea
where
\be
\xi_n \equiv \td c\, v^n \, \Big(\prod_{\a=1}^N Q_\a\Big)^{n\,v/2}\,\,
\sum_{\a=1}^N \fft{1}{(Q_\a)^n}\ . 
\ee
Note that the first two sub-leading terms depend on the charges only
through the function $\xi_1$, but that a new structure
$\xi_3$, with different functional dependence on the charges, emerges at
the third order in the expansion.  Similarly, yet another function of
the charges, $\xi_5$, emerges at the fifth order, and so on.  In
general, the expansion (\ref{st}) up to order $T^{m\,v}$ will involve
the $q$ independent functions $\{\xi_1,\xi_3,\xi_5,\ldots, 
\xi_{2q-1}\}$, where $q=[(m+1)/2]$ is the integer part of $(m+1)/2$.

    Having shown that the leading-order relationship between entropy
and temperature for near-extremal $p$-branes coincides with the
leading order in the relationship between the scattering cross-section
and the frequency for low-frequency scalar s-waves in the corresponding
extremal $p$-branes, it is of interest to investigate how far this
parallel extends at higer orders in the expansion.  There are only a
limited number of cases where higher-order corrections to the scattering
cross-sections have been reliably obtained.  One of these is the D3-brane,
for which the first few orders are \cite{gubhash}
\be
\fft{\sigma}{\sigma_0} = 1 -\ft16 Q\, \omega^4\,
 \log(e^{\gamma}\, Q^{1/4}\, \omega) +\ft{7}{72}\, Q\, \omega^4 
+ \cdots
\ee
where $\gamma$ is the Euler's constant.  On the other hand, the
relation between the entropy and temperature for the near-extremal
D3-brane can be seen from (\ref{st}) to be given by
\be
\fft{S}{S_0} = 1 + 4\pi^4\,  Q\, T^4 + \cdots\ .
\ee
Another example is the dyonic string in $D=6$, for which the first
few terms in the expansion for the s-wave scattering cross-section are
\cite{clpt}
\be
\fft{\sigma}{\sigma_0} = 1 - 2 (Q_e + Q_m)\,  \omega^2 \log(e^\gamma\,
\lambda) + (Q_e + Q_m)\, \omega^2 + \cdots\ .
\ee
Here we have defined $\lambda^2 = \sqrt{Q_e\, Q_m}\, \omega^2/4$.  By
contrast, the relation between the entropy and temperature, which can
be read off from (\ref{st}), is
\be
\fft{S}{S_0} = 1 + 8 \pi^2\, (Q_e + Q_m)\, T^2 + \cdots\ .
\ee
Thus we see that in these two examples there are additional terms,
depending upon the logarithm of the frequency, in the relation between
the scattering cross-section and the frequency.  

      A further example is the 3-charge black hole in $D=5$.  The low
energy absorption cross section of the extremal black hole was
obtained in \cite{mast}, in the case where the charge $Q_3$ is much
smaller than the other two charges $Q_1$ and $Q_2$.  It is given by
\be
\fft{\sigma}{\sigma_0} = 1 + \ft14 \pi\, \Big(\fft{Q_1\, Q_2}{Q_3}
\Big)^{1/2}\, 
\omega + \ft1{48}\pi^2\, \fft{Q_1\, Q_2}{Q_3}\, \omega^2
+ {\cal O}(\omega^4) \ .
\ee
The corresponding expression for the entropy in terms of the temperature
can be read off from (\ref{st}). For the case where $Q_3 << Q_1, Q_2$,
we find that it is given by
\be
\fft{S}{S_0} = 1 + \pi\, 
\Big(\fft{Q_1\, Q_2}{Q_3}\Big)^{1/2}\, T +
\ft32 \pi^2\, \fft{Q_1\, Q_2}{Q_3}\, T^2 +
2 \pi^3\, \Big(\fft{Q_1\, Q_2}{Q_3}\Big)^{3/2}\, T^3 +
{\cal O}(T^4)\ .
\ee
Thus we see that the expansion parameters are exactly the same,
although the precise numerical coefficients disagree.

\appendix

\section{Asymptotic forms of Bessel and Kummer functions}

   The asymptotic expansions of the Bessel functions for large values
of their arguments are
\bea
J_\nu(x) &\sim& \sqrt{\fft{2}{\pi\, x}}\, \cos(x-\ft12\nu\, \pi
-\ft14\pi)\ ,\nn\\
Y_\nu(x) &\sim&  \sqrt{\fft{2}{\pi\, x}}\, \sin(x-\ft12\nu\, \pi
-\ft14\pi)\ .\label{largex}
\eea
For small arguments, the Bessel functions can be approximated by
\bea
J_\nu(x) &=&  \fft{1}{\Gamma(\nu+1)}\, \Big(\fft{x}{2}\Big)^\nu -
 \fft{1}{\Gamma(\nu+2)}\, \Big(\fft{x}{2}\Big)^{\nu+2} +\cdots \
 ,\nn\\
Y_\nu(x) &=& - \fft{\Gamma(\nu)}{\pi}\,
\Big(\fft{x}{2}\Big)^{-\nu} +  \cdots\ .\label{smallx}
\eea

  Kummer's confluent hypergeometric functions $M(a,b,z)$ and
$U(a,b,z)$ satisfy the differential equation $z\, w'' + (b-z)\, w' -
a\, w=0$.  The solution $M(a,b,z)$ is known as the regular solution,
and it has the power-series expansion
\bea
M(a,b,z) &=& \sum_{n\ge 0} \fft{(a)_n\, z^n}{(b)_n\, n!} \nn\\
&=& 1 + \fft{a\, z}{b} + \fft{a(a+1)\, z^2}{2!\, b(b+1)}
+ \fft{a(a+1)(a+2)\, z^3}{3!\, b(b+1)(b+2)} +\cdots\ ,
\label{mseries}
\eea
where $(a)_n= a(a+1)(a+2)\cdots (a+n-1)$ is the ascending Pochhammer
symbol.  The irregular function $U(a,b,z)$ is expressible in terms of
$M(a,b,z)$ as follows:
\be
U(a,b,z) = \fft{\pi}{\sin\pi b}\, \Big[ \fft{M(a,b,z)}{\Gamma(1+a-b)\,
\Gamma(b)} - z^{1-b}\, \fft{M(1+a-b,2-b,z)}{\Gamma(a)\, \Gamma(2-b)}
\Big]\ .\label{udef}
\ee
This, together with (\ref{mseries}), can be used to determine the
small-$|z|$ dependence of the confluent hypergeometric functions.

   The asymptotic behaviour of the confluent hypergeometric functions
at large $|z|$ is as follows:
\bea
U(a,b,z) &\sim& z^{-a} \, \Big( 1 + O(z^{-1}) \Big)\ ,
\nn\\
M(a,b,z) &\sim& \fft{\Gamma(b)\, e^{\pm\im\pi a}\,
z^{-a}}{\Gamma(b-a)}\,\Big(1 +
O(z^{-1}) \Big)
+ \fft{\Gamma(b)\, e^z\, z^{a-b} }{\Gamma(a)} \Big(1 + O(z^{-1})
\Big)\ ,\label{muasymp}
\eea
where in the second line the upper sign is taken if $-\ft12\pi < {\rm
arg}\, z < \ft32\pi$, whilst the lower sign is taken if $-\ft32\pi <
{\rm arg}\, z \le -\ft12\pi$.

\end{document}